\documentclass[twocolumn,showpacs,preprintnumbers,amsmath,amssymb,aps,pre]{revtex4}
\usepackage{graphicx,color}
\newcommand{\EQ}{\begin{equation}}
\newcommand{\EN}{\end{equation}}
\newcommand{\EQA}{\begin{eqnarray}}
\newcommand{\ENA}{\end{eqnarray}}

\begin{document}
\title{Anomalous scaling and generic structure 
function in turbulence}
\author{ B. Dubrulle$^{1,2}$ }

\affiliation{$^1$ CNRS, UPR 182, CEA/DSM/DAPNIA/Service d'Astrophysique, CE Saclay,  F-91191 Gif sur Yvette, France\\
$^2$ CNRS, URA 285, Observatoire Midi-Pyr\'en\'ees, 14 avenue Belin, F-31400 Toulouse, France\\}

\begin{abstract}

We discuss on an example a general mechanism of apparition of anomalous
scaling in scale invariant systems via zero modes of a scale invariant 
operator. We discuss the relevance of such mechanism in turbulence, and
point out a peculiarity of turbulent flows, due to the existence of
both forcing and dissipation. Following these considerations, we 
show that if this mechanism of anomalous scaling is operating in 
turbulence,
the structure functions can be constructed by simple 
symmetry considerations.
We find that the generical scale behavior of structure functions 
in the inertial range is not self-similar $S_n(\ell)\propto
\ell^{\zeta_n}$ but includes an "exponential self-similar" 
behavior $S_n(\ell)
\propto \exp[\zeta_n\alpha^{-1} \ell^{\alpha}]$ where $\alpha$ 
is a parameter proportional to the inverse of the logarithm of 
the Reynolds number. 
The solution also follows exact General Scaling and 
approximate Extended Self-Similarity.

\end{abstract}


\date{ Submitted to Journal de Physique France: April 96- Revised: August 96}

\maketitle

\begin{center}
{\bf R\'esum\'e}
\end{center}

A partir d'un exemple, nous discutons un m\'ecanisme g\'en\'eral 
de production de lois d'\'echelle anormales, dans un syst\`eme 
invariant d'\'echelle. Ce m\'ecanisme repose sur l'existence de valeurs
propres nulles pour un op\'erateur invariant d'\'echelle. Nous 
discutons ensuite la pertinence de ce m\'ecanisme en turbulence,
en soulignant une particularit\'e des \'ecoulements turbulents, 
li\'ee \`a la coexistence d'un for\c cage et de la dissipation. En
utilisant ces consid\'erations, nous montrons que si ce m\'ecanisme
s'applique \`a la turbulence, alors on peut construire les fonctions
de structure par de simples arguments de sym\'etrie. On trouve que
le comportement g\'en\'erique des fonctions de structure n'est pas 
auto-similaire $S_n(\ell)\propto \ell^{\zeta_n}$ mais inclut un 
terme "exponentiel auto-similaire" du type $S_n(\ell)
\propto \exp[\zeta_n\alpha^{-1} \ell^{\alpha}]$, o\`u $\alpha$ est 
un param\`etre inversement proportionnel au logarithme du nombre
de Reynolds. La solution satisfait \'egalement rigoureusement la 
propri\'et\'e de Similarit\'e G\'en\'erale et approximativement 
la propri\'et\'e d'Autosimilarit\'e Etendue.

\section{Introduction}

Anomalous scaling usually refers to situations where the observed 
scaling exponents deviate from their natural dimensional scaling. 
As a consequence, it is often associated with non-simple scaling
behavior of the correlation functions: the scaling exponent 
of the 2n$^{th}$ order correlation function is not $n$ times the
scaling exponent of the 2$^{nd}$ order correlation function. A good
example of this situation is the scaling of the so-called velocity
structure functions in 3D homogeneous isotropic turbulence. 
Dimensional
considerations based on energy conservation 
\cite{K41} predict that the scaling
exponent of the $n^{th}$ structure function should be $\zeta_n=n/3$.
Observed scaling exponents substantially deviate from this 
prediction (see e.g. \cite{arneodo} for a review of recent 
 experimental results). Several general mechanisms of apparition 
of anomalous scaling have been identified. Among them, two 
mechanisms deserve special interest: existence of
additional essential parameters which cannot be eliminated 
from the problem (this is the phenomenon of incomplete
self-similarity, reviewed by \cite{barenblatt}) and existence 
of non-trivial zero modes of the closed equations satisfied by the 
correlation functions \cite{zeromodes}.\

Our aim is to study some consequences of 
these two form of anomalous scaling in situations where
the system is scale invariant. Specifically, we show on a 
simple example how zero-modes give rise to a self-similar behavior
of second kind. We then come back to the
case of turbulence to show how this feature can be used to determine
general properties of the scaling exponents and the scale behavior
of the structure functions. The main conclusion of the present 
study is that the generic scale behavior of structure functions 
in the inertial range is not self-similar $S_n(\ell)\propto
\ell^{\zeta_n}$ but "exponential self-similar" $S_n(\ell)
\propto \exp[\zeta_n\alpha^{-1} \ell^\alpha]$ where $\alpha$ 
is a parameter proportional to the inverse of the logarithm of 
the Reynolds number. 
The self-similar shape is recovered only in the limit
of infinite size inertial range (infinite Reynolds number). 

\section{Anomalous scaling in scale invariant systems}

The goal of the present Section is not to prove rigorous 
results, but to illustrate a possible mechanism
of generation of anomalous scaling in turbulence, and explore 
its consequences. In particular, the discussion about application 
to turbulence should be taken as rather speculative, since 
we do not derive any exact equations. The main 
goal of the present Section is to give justifications for the 
symmetry approach presented in the next Section, which will
provide some new and quantitative results on structure functions
in turbulence.

\subsection{An example}

Consider the differential equation:
\EQ
\partial_t \phi+\partial_\ell \phi-\frac{\zeta}{\ell}\phi=F(\ell),
\label{modele}
\EN
where $t$ is the time, 
$\ell$ is the scale, $\phi$ is a scale-dependent 
physical quantity (e.g. a structure function, or
a correlation function), $F$ an external forcing and $\zeta$ a 
constant. We further assume that the scale range is bounded 
by some larger scale $L$, at which $\phi$ obeys the boundary 
condition $\phi(\ell=L)=0$, and that, for scales much smaller than
the larger scale, $F$ follows a power-law $F(\ell)\propto\ell^{h_0}$.
This property can be obtained for example if $F$ is a Gaussian random function
with a power-law spectrum.

\subsubsection{Scale invariance}

We note that, in absence of forcing, ($F=0$), the equation 
(\ref{modele}) is invariant under the family of transformations 
${\cal S}_h(\lambda)$:
\EQ
{\cal S}_h(\lambda):\;\ell\to \lambda \ell,\quad \phi\to \lambda^h \phi,
\quad t\to \lambda t,
\label{scalesym}
\EN
for arbitrary $\lambda$ and $h$. The physical 
system described by eq. (\ref{modele}) without forcing
is then said to be scale invariant. Several important properties
of scale invariant systems are listed in Section 3.1. 

\subsubsection{Zero-mode}

In absence of forcing, stationary solutions of eq. (\ref{modele})
 satisfy the homogeneous equation:
\EQ
\partial_\ell \phi-\frac{\zeta}{\ell}\phi=0.
\label{modelesf}
\EN
The solution is $\phi=A\ell^\zeta$, where $A$ is to be chosen according
to the boundary condition, namely: $\phi(L)=AL^\zeta=0$, i.e. 
$A=0$. In absence of forcing, there are therefore no non trivial 
stationary solutions of eq. (\ref{modele}). The situation is 
modified when the forcing is taken into account.

\subsubsection{Dimensional analysis}

Before solving the forced equation, let us perform a simple dimensional
analysis of eq. (\ref{modele}). We note that for scale smaller than
$L$, stationary solutions can be found by using the scaling 
properties of $F$ (Section 2.1) and balancing the l.h.s. and
the r.h.s. of (\ref{modele}):
\EQ
\frac{\phi}{\ell}\sim F\sim \ell^{h_0},\; \ell\ll L,
\label{dimension}
\EN
which predicts another power-law solution $\phi\propto \ell^{1+h_0}$,
for $\ell\ll L$. 

\subsubsection{Incomplete similarity}

The previous dimensional analysis ignores the influence of the largest
scale $L$. In fact, the solution must be influenced by the scale $L$,
would it be only to satisfy the boundary condition $\phi(L)=0$
(which is not satisfied if the solution is only a power law!). A 
correct dimensional analysis thus imposes:
\EQ
\phi(\ell)=\ell F(\ell) G(\ell/L),
\label{plus}
\EN
where $G$ is a function which cannot be determined by dimensional
arguments. A modification of the dimensional power law
solution $\phi\propto \ell^{1+h_0}$ can then be observed if the 
variable
$L$ is essential and if $G$ obeys incomplete similarity in $L$, i.e.:
\EQ
G(\ell/L)\sim (\ell/L)^{-\Delta},\; L\gg 1,\; \Delta>0.
\label{essential}
\EN
Since $\Delta$ is positive, the function $G$ tends to infinity when 
$L$ tends to infinity. In other words, the size of the system is 
always relevant and cannot be eliminated from the problem. 
In such a case, $\phi$ follows an {\sl anomalous scaling}:
$\phi\propto L^\Delta \ell^{1+h_0-\Delta}$. The value of $\Delta$
can be found by solving exactly the equation (\ref{modele}).

\subsubsection{Exact solution of the complete problem}

The stationary solution of eq. (\ref{modele}) is:
\EQ
\phi=A\ell^\zeta+\ell^\zeta\int_0^\ell \rho^{-\zeta}F(\rho)d\rho.
\label{exactsol}
\EN
The constant $A$ is found by imposing the boundary condition at 
$\ell=L$:
\EQ
A=-\int_0^L\rho^{-\zeta}F(\rho) d\rho\sim L^{1+h_0-\zeta}.
\label{constante}
\EN
The leading contribution to $\phi$ for $\ell/L\ll 1$ 
then depends on the sign 
of $\Delta=1+h_0-\zeta$. When it is negative, $\phi$ just obeys 
the dimensional scaling, and the contribution due to $L$
tends to zero as $L\to\infty$. This is the case of complete 
similarity, with no anomalous scaling. When $\Delta$ is positive,
the dominant contribution is $L^\Delta \ell^{\zeta}$, 
which tends to infinity with $L$. This is the case of incomplete 
similarity described previously.

\subsubsection{Summary}

This simple example illustrates the general mechanism of anomalous 
scaling discussed in \cite{zeromodes} in the case 
of a finite size, scale-invariant system: 
a zero mode, with a generic
non-dimensional similarity exponent, and which does not satisfy 
the boundary condition, becomes relevant in the presence of an 
external forcing, which also imposes a dimensional, normal scaling
exponent. The dominant contribution to the solution then depends
on the relative value of the two scaling exponents, and in particular,
on the behavior of the limit $L\to\infty$.

\subsubsection{Implications on scaling exponents}

In the case discussed previously, non-anomalous exponents are 
given by the exponents of the zero mode of a scale invariant differential
operator. It then makes a lot of sense to investigate the possible
constraints on scaling exponents dictated by the scale symmetry.
This fully justifies computation of scaling exponents based 
on multi-fractal theory \cite{mfr}, conformal theory 
\cite{polyakov}, or more recently on a principle of exponent
relativity developed in \cite{DG1,pocheau2}. Such approaches 
restrict more or less 
severely the possible shape of non-anomalous scaling in scale invariant
systems. Restrictions are very weak in the case of the multi-fractal
or conformal theory, while they are rather strong in the exponent
 relativity approach:
in that case, the possible generic exponents fall into three classes
depending on the properties of the rarest events. Within each class,
the shape of the scaling exponents as a function of the order of
the correlation function is entirely determined by the scale symmetry,
and only depends on three parameters.\

Transition from anomalous to non-anomalous scaling can also be 
constrained within the present model. Obviously, the value of 
the scaling exponents cannot be affected by modifications of 
external parameters, such as the strength of the force. The transition
to non-anomalous scaling can only be obtained by a modification of 
the shape of the differential operator, i.e. by variation of 
internal parameters such as dimensionality, or isotropy properties.
In the case of the white advected passive scalar, for example, 
it has been proved \cite{zeromodesbis}
that the non-anomalous behavior disappears 
in the limit where the (topological) 
dimension of the system is infinite. This kind of asymptotic 
behavior could well be found in a large variety of systems, such as
turbulence.

\subsection{The case of turbulence}

The Navier-Stokes equations are:
\EQ
\partial_t {\vec u} + ({\vec u}\cdot{\vec \nabla}) {\vec u} + 
{\vec \nabla} P =
\nu \nabla^2 {\vec u} + {\vec F},\quad {\vec \nabla} \cdot 
{\vec u} =0,
\label{navier}
\EN
where $t$ is the time, ${\vec u}$ the velocity
field, $P$ the pressure, $\nu$ the viscosity and ${\vec F}$ the external 
force, acting at large scales.
Turbulence is defined as a statistically stationary
regime of the Navier-Stokes equations where an external forcing
balances the viscous energy losses and induces a constant energy
rate: $du^2/dt=cte$. This means that the product ${\vec F}\cdot {\vec u}$
is scale independent, so that the force 
scales like $1/u$. We note that eq. (\ref{navier}) exhibits common
features with eq. (\ref{modele}). The l.h.s. of (\ref{navier}) is
invariant under the family of scale dilations
${\cal S}_h({\lambda})$ with arbitrary similarity exponent $h$
and scale factor ${\lambda}$ 
\cite{frischbook}:
\EQ
{\cal S}_h({\lambda}): \quad
{\ell} \to {\lambda} {\ell},\quad u\to {\lambda}^h u,
\quad t\to {\lambda}^{1-h}t,
\label{scaleinva}
\EN
where $\ell$ is the space scale. This scale symmetry property is 
broken by both the forcing and the dissipation, which considered 
separately impose a dimensional scaling exponent: for the 
forcing, it is obtained by balancing $F$ and $du/dt$, which
given the transformation properties of $F$, yields $u^3\sim \ell$,
i.e. $h_F=1/3$. For the dissipation, balancing of $du/dt$ with
$\nu \nabla^2 u$ gives $u\sim \ell^{-1}$, i.e. $h_D=-1$.\

Experimentally, none of these dimensional scalings is observed.
Let us define the longitudinal velocity increments
at scale $\ell$ by $\delta u_\ell={\vec \ell}\cdot [{\vec u}({\vec
x}+{\vec \ell})-{\vec u}({\vec
x})]/\ell$ and the velocity structure function of order $n$, $
S_n(\ell)=<(\delta u_\ell)^n>$, where $<>$ is a statistical average.
In a range of scales small compared with the forcing scale, and 
large compared with the scales at which dissipation becomes 
non-negligible (the so called inertial range), the structure functions
are observed to follow approximate power law behavior:
$S_n(\ell)\propto\ell^{\zeta_n}$. The dimensional value imposed by the
forcing would be 
$\zeta_n=nh_F=n/3$ while the dimensional value imposed by 
dissipation would be $\zeta_n=nh_D=-n$. Experimentally, none of 
these dimensional scaling is observed. The origin of this anomalous scaling
can be understood within the mechanism explored previously: following
the Navier-Stokes equation, the structure functions obey an equation
of the type:
\EQ
I_n S_n=F_n+D_n,
\label{type}
\EN
where $I_n$ is a scale invariant differential operator acting in the 
scale space, and $F_n$ and $D_n$ contributions due to forcing and 
dissipation, imposing dimensional exponents $n/3$ and $-n$. The
scaling exponents in the inertial range then come from the scaling
properties of the zero-mode of the operator $I_n$. 
Note that until now, the explicit shape of the operator $I_n$
has never been computed, even perturbatively. It is quite possible
that it is a non-diagonal operator of the structure functions, like 
in the Burgers equation, in which case the explicit computation of
the zero-modes would be very difficult. Note also that for the 
Burgers equation, the scaling comes from the balance of non-diagonal 
contributions with terms
corresponding to $D_n$, which shows that the anomalous scaling described
in the previous Section does not necessarily apply for any type of
equation. In this respect, our model equation (\ref{modele}) is a 
little bit ambiguous, and its application to turbulence not so 
straightforward as it may seem. We however adopt it as a working 
hypothesis, to explore its consequences.\ 

Because of the presence of two scale-breaking symmetry mechanisms,
turbulence offers a new possibility absent in the 
model equation (\ref{modele}): transition to new anomalous 
exponents at small scales, obtained by balancing of $F$ and $D$.
Because of constant energy transfers, $F$ scales like $1/u$, 
while $D$ scales like $u/\ell^2$, so 
$h_T=1$. Indeed, experimentally, one observes a transition towards
$S_n\propto \ell^n$ at small scales, the so-called ``regular
behavior". The existence of this transition imposes a scale 
dependence for the scaling exponents, and thus, non trivial scale
dependence for the shape of the structure functions. Of course,
this shape depends both on the shape of the operator $I_n$, and 
on $F_n$ and $D_n$. Given the symmetry properties of $I_n$, 
it is however interesting to investigate whether the scale 
symmetry does impose possible generic shape for 
the structure functions, even within the inertial range, in the same 
way as it does constrain the $n$ dependence of the scaling 
exponents. This is the purpose of the next Section.

\section{Scale covariant structure functions in turbulence}

In the present Section, we assume that the velocity structure functions
in turbulence obey a differential equation which can be put 
under the form (\ref{type}), where $I_n$ is a scale invariant
operator. Furthermore, we assume that the solution in the inertial
range is given by the homogeneous part of the equation 
($F_n=D_n=0$), and that there is a transition towards the "regular"
solution $S_n(\ell)=K_n \ell^n$ at a scale $\ell=\eta_n$, 
where both $K_n$ and $\eta_n$ are entirely 
determined by $F_n$ and $D_n$. Our 
inertial range solution therefore extends up to $\eta_n$. Our aim is to derive
a generic shape for the structure functions, satisfying all these 
requirements. From now on, we focus on stationary situations, 
thereby neglecting any time dependence.

\subsection{On scale symmetry}

A scale invariant system is according to our definition, 
invariant by a transformation ${\cal S}_h(\lambda)$ given 
in (\ref{scalesym}). Note that in log-variables $A=\ln\phi$, $T=\ln\ell$,
this transformation amounts to a translation invariance in $A$ and 
$T$. This symmetry is rather intuitive, and connected with the 
possibility to multiply any characteristic quantity by an 
arbitrary constant, i.e. to 
perform arbitrary changes of units in the system
\cite{DG1}. This corresponds to
a global scale invariance. In fact, as discussed by Pocheau
\cite{pocheau}, scale symmetry encompasses another notion, corresponding
to symmetry with respect to local scale dilations. This 
symmetry is connected with the invariance of the system with respect
to changes of resolution, or equivalently to the possibility
of performing arbitrary changes of rationalized (power law) unit 
systems \cite{DG1}. To understand this local symmetry, it is convenient
to consider a discrete slicing of the space scale $\ell\le L$, 
under the form:
\EQA
\ell_i&=&\Gamma^i\ell_0,\quad \phi_i=\Lambda^i \phi_0,\nonumber\\
L&=&\Gamma^N \ell_0.
\label{discrete}
\ENA
Here, $\Gamma$ (or equivalently $N$) and $\Lambda$ characterize the 
resolution of the 
observations. By local scale symmetry, we require that the 
system be invariant by changes of resolution, i.e. by changes
of $N$. Changes in $N$ are achieved by dilations of 
$\Gamma$: $\Gamma\to \Gamma^\alpha$, which transform $i$ into 
$i\alpha$. It is also a transformation which transform 
$\ell_i/\ell_j$ into $(\ell_i/\ell_j)^\alpha$, and 
$\phi_i/\phi_j$ into $(\phi_i/\phi_j)^\alpha$. A local scale dilation
can therefore be seen as a transformation:
\EQ
\ell/\ell_0\to (\ell/\ell_0)^\alpha,\quad \phi/\phi_0\to 
(\phi/\phi_0)^\alpha,
\label{local}
\EN
for arbitrary $\alpha$, 
$\ell_0$ and $\phi_0$. It can be seen as a change of rationalized 
unit system because it amounts to change $\ell_0$ and $\phi_0$ 
by $\ell_0 (\ell/\ell_0)^{1-\alpha}$ and 
$\phi_0 (\phi/\phi_0)^{1-\alpha}$.\

Summarizing, we say that a system is scale invariant, if it is invariant
under both (\ref{scalesym}) and (\ref{local}), i.e. if the 
equations
governing its evolution are covariant (keep the 
same shape) under these two transformations.

\subsection{Generic equation for structure function}

Assuming that a turbulent flow is scale invariant, we must then write 
the equations followed by a structure function $S_n(\ell)$
in a covariant 
way. For convenience, we introduce the log variables 
$X_n=\ln S_n$ and $T=\ln\ell$ \cite{remark0}. 
The shape of the structure function is
 determined by the function $X_n(T)$, given some boundary (initial)
conditions. We must write the differential equation followed 
by $X_n$. Because 
of the global symmetry (\ref{scalesym}), which amounts to a translation
invariance in $X_n$ and $T$, this differential equation can only 
include derivatives of $X_n$ with respect to $T$. In the spirit of
the amplitude equation theory, we  write the generic equation
as an expansion in power of the amplitude:
\EQA
\chi_n&=&a_n\frac{dX_n}{dT}+b_n \frac{d^2 X_n}{dT^2}+c_n\frac{d^3 X_n}{dT^3}+
O(\frac{d^4 X_n}{dT^4})\nonumber\\
&+&
g_n\biggl(\frac{dX_n}{dT}\biggr)^2+h_n\frac{d X_n}{dT}\frac{d^2 X_n}{dT^2}+
O((\frac{d X_n}{dT})^3),
\label{ampli}
\ENA
where $\chi_n$,$a_n$, $b_n$, ...$h_n$ are some $T$ independent 
constants characterizing the system
and are to be chosen in order to respect the local scale symmetry.
 Applying the transformation (\ref{local}) to (\ref{ampli}), we see
that this equation is covariant only if the constants satisfy the 
following transformation rule:
\EQA
&&\chi_n\to \chi_n;\;a_n\to a_n;\;\; b_n\to \alpha b_n; c_n\to \alpha^2 c_n;
\nonumber\\
&& g_n\to g_n;\; h_n\to \alpha h_n; ...
\label{conditions}
\ENA
This suggests to introduce two characteristic constant 
scales in the system, 
$\ell_0$ and $\ell_1$ and to write:
\EQA
\chi_n&=&{\tilde \chi_n},\nonumber\\
a_n&=&{\tilde a_n},\nonumber\\
b_n&=&{\tilde b_n} \ln(\ell_1/\ell_0),\nonumber\\
c_n&=&{\tilde c_n} (\ln(\ell_1/\ell_0))^2,\nonumber\\
g_n&=&{\tilde g_n},\nonumber\\
h_n&=&{\tilde h_n} \ln(\ell_1/\ell_0),
\label{suggests}
\ENA
where the constants ${\tilde \chi_n},...$ are invariant under the local scale 
transformation, and only depend on the structure function (i.e. on $n$) 
and on the choice of $\ell_0$ and 
$\ell_1$. From now on, we fix the scale origin at 
the integral scale $\ell=L$ (representing the largest correlated motions). 
With this choice, scale invariance
symmetry holds for 
\EQ
-\ln(R_*)\le T\le 0,
\label{scaleihol}
\EN
where $R_*=L/\eta$ is a dimensionless number apparented to a Reynolds number
and $\eta$ is. 
the Kolmogorov scale (representing dissipative scales of turbulence). 

\subsection{The case of turbulence}

Equation (\ref{ampli}), with constants following (\ref{suggests}) 
is a generic equation in a scale invariant system, satisfying 
both global and local symmetry. Its solution depends on the 
values of the constants, which must be computed following some
systematic procedure. 
However, interesting features of the solution
may already be found in the linear approximation, where all terms
of degree $A^2$ or higher are neglected
\cite{remark2}. This approximation may be justified by considering 
a range of scales where the amplitude $A$ is small enough.
We study here only the simplest case relevant to
turbulence.

\subsubsection{The linear case}

The highest relevant derivative is determined by the number of 
boundary conditions. The existence of a largest scale in the 
system (e.g. the injection scale) imposes already at least 
one boundary condition. Furthermore, the existence of
the transition between the scale invariant
solution and the regular solution at $\ell=\eta_n$ imposes at least
a second boundary condition. Note that this is a peculiarity of
turbulence. In many scale invariant systems, only one boundary 
condition is needed, corresponding to an infrared or ultraviolet
 cut-off in the scale space.\

\begin{figure}[!htb]
\includegraphics[clip=true,width=0.99\columnwidth]{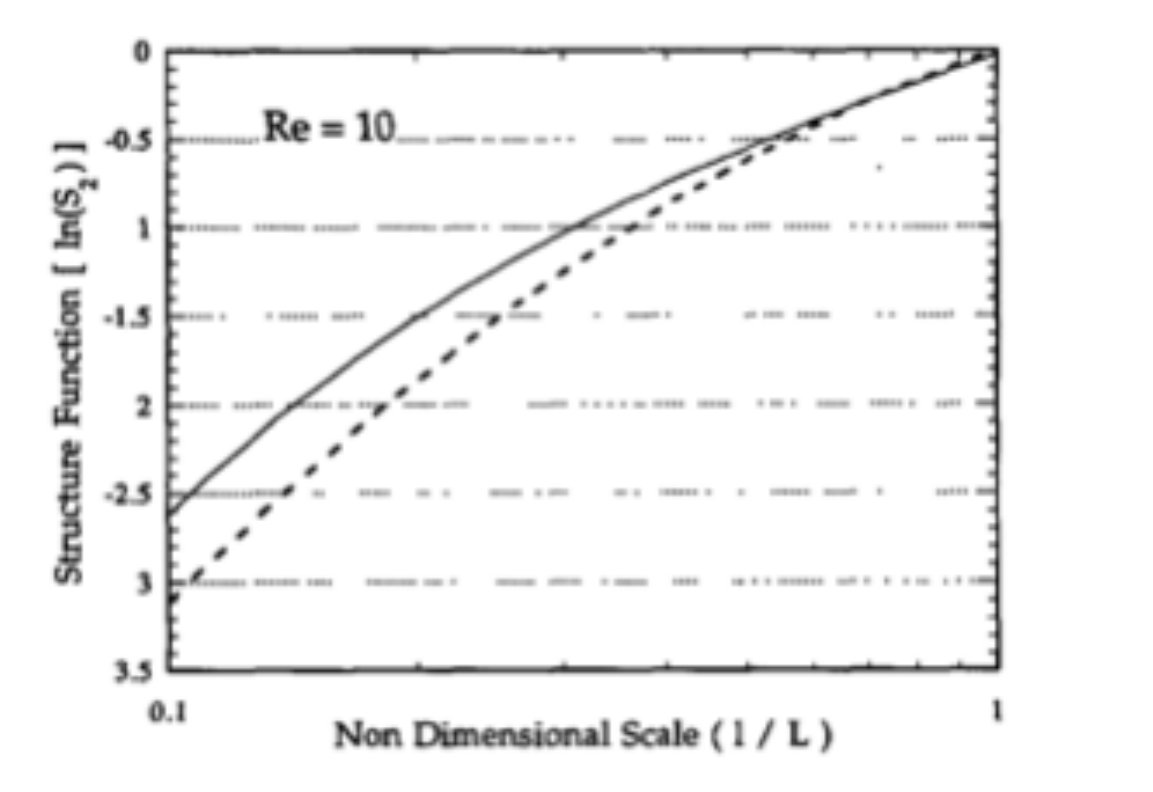}
\caption{The behavior of a structure function versus the scale when 
$a_n=0$ (dotted line) and 
$a_n\neq 0$ (continuous line).}
\label{fig:one}
\end{figure}

We adopt $L$, the integral scale 
scale, and $\eta_n$  to normalize the 
constants, and introduce the pseudo-Reynolds number
$$\ln(R_n)=\ln(L/\eta_n). $$ Dropping tildes for convenience,
we therefore write
the simplest differential equation relevant to structure 
functions in turbulence 
as:
\EQ
\chi_n=a_n\frac{d\ln(S_n)}{dT}
+b_n\ln(R_n)\frac{d^2\ln(S_n)}{dT^2},
\label{casoli}
\EN
where $\chi_n$, $a_n$ and $b_n\neq 0$ are $T$ independent 
constant depending on the order 
of the structure function. Note that they also are independent on the 
Reynolds number, since solutions with different Reynolds number $R_1$ 
and $R_2$ can be 
related by a local scale transformation (\ref{local}) with 
$\alpha=\ln(R_1)/\ln(R_2)$.
 The solution is:
\EQA
\ln(S_n)&=&\frac{\chi_n}{2b_n\ln(R_n)}T^2+\alpha_nT+\beta_n,\quad a_n=0,
\nonumber\\
\ln(S_n)&=&\frac{\chi_n}{a_n}T+\alpha_n
e^{-(a_n T)/(\ln(R_n) b_n)}+\beta_n,\quad a_n\neq 0.
\label{solution}
\ENA
The constant $\alpha_n$ and $\beta_n$ are fixed by the boundary 
conditions. Without loss of generality, we can fix the scale 
origin at $\ell=L$, and choose the normalization for $S_n$ 
such that:
\EQA
\ln(S_n)(T=0)&=&0,\nonumber\\
\frac{d\ln(S_n)}{dT}\vert_{T=-\ln(R_n)}&=&n.
\label{boundary}
\ENA
The second boundary condition guarantees the transition towards 
the regular solution. These boundary conditions fix the constants as:
\EQA
\alpha_n&=&n+\frac{\chi_n}{b_n};\;\beta_n=0,\quad a_n=0,
\nonumber\\
\alpha_n&=&-\beta_n=\biggl(\frac{\chi_n}{a_n}-n\biggr)
\frac{b_n \ln(R_n)}{a_n}e^{-a_n/b_n},\quad a_n\neq 0.
\label{solucon}
\ENA
The solutions (\ref{solution}) with constant given by (\ref{solucon})
both follow the behavior schematized in Fig. \ref{fig:one}: a power law 
behavior around $\ell=L$, followed by a transition regime towards
$\ell=\eta_n$, where matching with the regular solution is obtained.\

The second solution, with $a_n\neq 0$ 
(generic case) is the most interesting,
because it exhibits a feature
reminiscent of what is observed in turbulence. This is the subject of
the remaining Sections.

\subsubsection{Large Reynolds number limit}
 
In the limit where the pseudo-Reynolds number is very large, 
$T/\ln(R_n)\ll 1$ and the argument of the exponential can 
be expanded. 
The generic solution then approaches the self-similar behavior:
\EQ
\ln(S_n)=\biggl(\frac{\chi_n}{a_n}(1-e^{-a_n/b_n})+n
e^{-a_n/b_n})\biggr)
\ln(\ell/L).
\label{limite}
\EN
This behavior is also valid in the neighborhood $\ell=L$.
This kind of asymptotic behavior has been observed in a tunnel wind 
experiment by Castaing et al \cite{castaingetal}: when the Reynolds number 
is increased, the shape of the structure function goes from 
exponential power law, like in the first part of our generic solution
(\ref{solution}b), to self-similar power law behavior. Even in the 
moderate Reynolds number regime, Castaing and his collaborators 
observe that the exponential power law regime is superseeded at 
larger scales by a pure power law regime. Such kind of behavior 
is indeed obtained at $\ell>L$ 
in the generic solution, provided $a_n/b_n$ is 
negative. This exponential power law as well as the variation of the 
exponent with ($1/\ln R$) has also been predicted by 
Castaing \cite{castaing} using a scale invariant Lagrangian 
formalism and requirement of finite dissipation.
Within the present framework the result appears to stem both from
scale invariance and finite size effects (see discussion). 

\subsubsection{Interpretation of the constants}

Up to this stage, we have not tried to characterize the various 
constants $\chi_n$, $a_n$ and $b_n$ appearing in the solution.
They should in principle be  directly computed from the Navier-Stokes
equations. Given the properties of the generic solution, we can
however propose an interpretation of the constants which provides
some constraints on their shape. Consider the asymptotic behavior
of the solution (\ref{limite}). It defines some  
"inertial range" scaling exponents:
\EQ
\zeta_n=
ne^{-a_n/b_n}
+\frac{\chi_n}{a_n}(1-e^{-a_n/b_n}).
\label{exponents}
\EN
They are made from two contributions: one stemming from the 
solution of the equation $dA/dT=\chi_n/a_n$; another one
coming from the exponential power term, present only when 
$b_n$ is different from zero, i.e. when the system is of second order.
As discussed in Section 3, the presence of this term is 
directly linked with the existence of a small scale cut-off, where
the solution must bifurcate towards the regular solution. It can 
therefore be interpreted as a finite-size effect. On the other hand,
the contribution at $b_n=0$ characterizes a perfect scale invariant 
solution, extending from $0$ to infinity in the scale space. This
separation is reminiscent of the model of Dubrulle and Graner
\cite{DG1}, in which generic scaling exponents can be written:
\EQ
\zeta_n=n\Delta_\infty+\delta\zeta_n.
\label{dgpred}
\EN
Here, $\delta\zeta_n$ is a contribution stemming only from scale 
symmetry considerations. It is characterized by its large $n$ limit
$C=\lim_{n\to \infty}\delta\zeta_n$ which can be interpreted as the codimension
of the most intermittent structures.
The term proportional to $\Delta_\infty$
is the contribution due to the scaling properties of the maximum value
of $\ln(S_n)$, which is defined only for finite-size systems. The
factorization (\ref{dgpred}) with a linear dependence of the 
contribution due to finite-size effect can also be justified from
dynamical considerations \cite{ugalde}. This suggests to interpret
the constants appearing in (\ref{exponents}) as:
\EQA
e^{-a_n/b_n}&\equiv&\Delta_\infty,\nonumber\\
\frac{\chi_n}{a_n}&\equiv&\frac{\delta \zeta_n}{1-\Delta_\infty}.
\label{interpretation}
\ENA
With this interpretation, finite size effects modifies the scale invariant
solution in two ways: by the introduction of the linear term $n\Delta_\infty$
and by the modification of the codimension of the most intermittent structure:
\EQ
C_{scale}=\frac{C_{finite}}{1-\Delta_\infty},
\label{modifi}
\EN
where $C_{scale}$ is the codimension in the scale invariant case, and 
$C_{finite}$ is the codimension in the finite size case. We stress that 
this finite size effect is Reynolds number independent (it only depends
on the value of the constant $b_n$). It may however depend on 
other external parameter, such as the dimension (see discussion of 
Section 2.1).\

Using eq. (\ref{solucon}), interpretation (\ref{modifi})
 completely determines the ratio
$\chi_n/a_n$ and $a_n/b_n$ as a function of $\Delta_\infty$ and
the function $\delta\zeta_n$, which can be severly constrained 
using only scale symmetry requirements \cite{DG1}. 

\subsubsection{Example: log-Poisson case}

\begin{figure}[!htb]
\includegraphics[width=0.99\columnwidth]{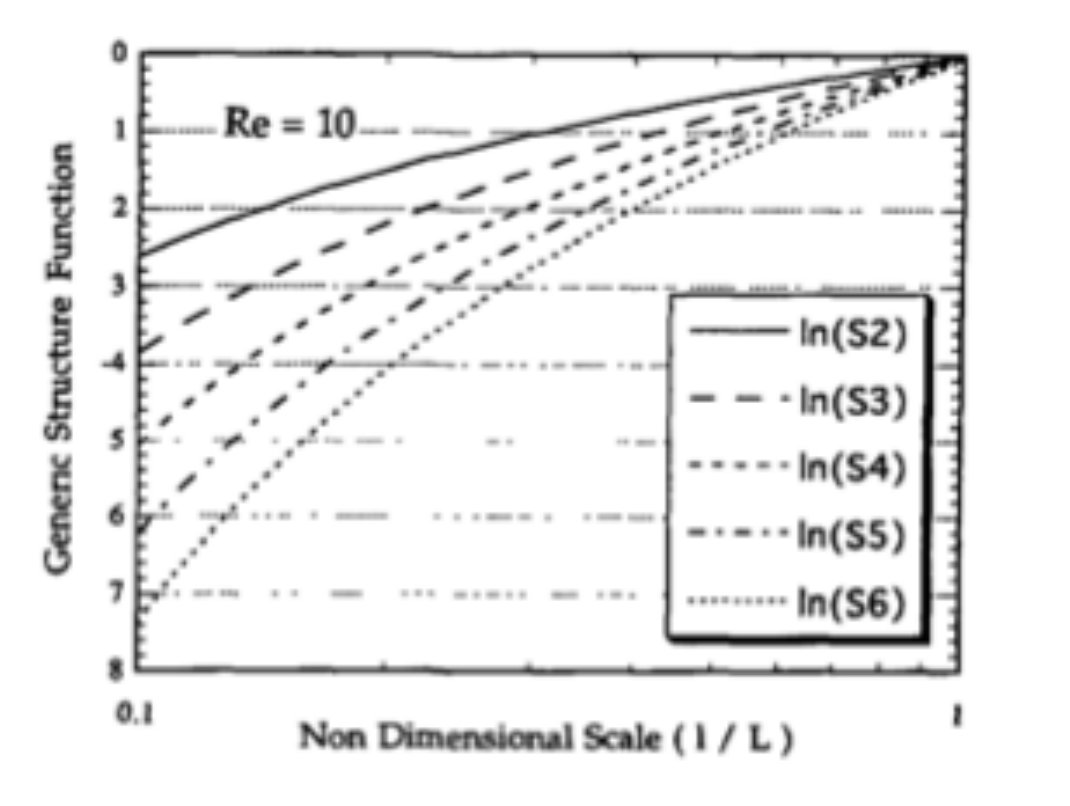}
\caption{The generic structure functions $S_2$ to $S_6$ in the log-Poisson 
case.}
\label{fig:two}
\end{figure}

As an example,
let us considered the case where $\delta\zeta_n$ corresponds
to a log-Poisson statistics 
\cite{SL,dubrulle,SW,DG2} (one of the three possible
cases derived in \cite{DG1}):
\EQ
\delta\zeta_n=C(1-\beta^n),
\label{poisson}
\EN
where $C$ and $\beta$ are two constants. If we adopt the values 
$C_{finite}=2$, $\beta^3=2/3$ and $\Delta_\infty=1/9$ as advocated by She and
Leveque \cite{SL}, for high Reynolds number turbulence, we can numerically 
determine the values of all parameters and thus, the scale
dependence of the structure functions, providing the 
pseudo-Reynolds number $R_n$ are given. Table \ref{table:one} summarizes the values
obtained for $n=2$ to $6$. The corresponding structure functions,
for $R_n=10$ are given in Fig. \ref{fig:two}.

\begin{table}
\begin {tabular}{|c|c|c|} \hline
$n$&$\chi_n/a_n$&
$a_n/b_n$\\ \hline
2&0.53&2.2\\
3&0.75&2.2\\
4&0.95&2.2\\
5&1.1&2.2\\
6&1.25&2.2\\
\hline
\end{tabular}
\caption{Values of the parameters appearing in the structure 
functions for the log-Poisson model.}
\label{table:one}
\end{table}

\subsubsection{Comparison with Batchelor's parametrization}

\begin{figure}[!htb]
\includegraphics[width=0.99\columnwidth]{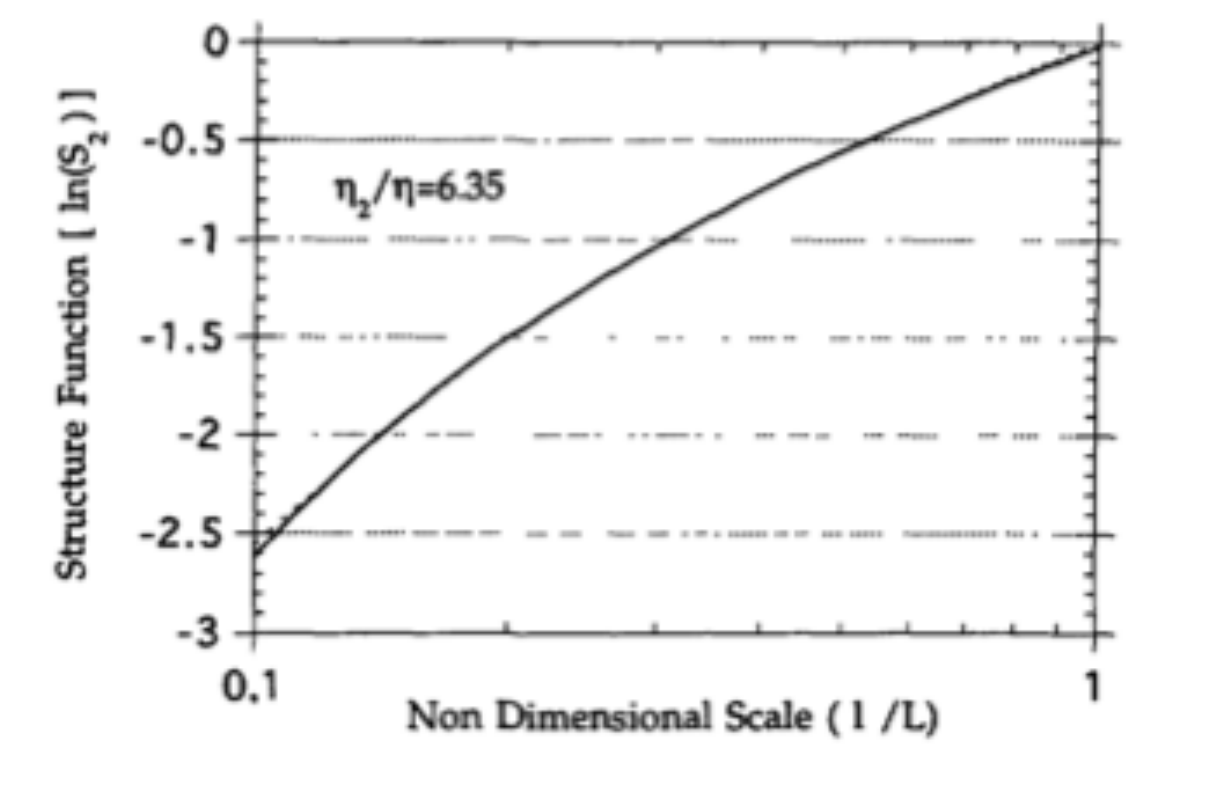}
\caption{Comparison between the Batchelor parametrization (dotted line) and 
the generic structure function $\ln(S_2)$ 
in the log-Poisson case (continuous line).}
\label{fig:three}
\end{figure}

Using only symmetry arguments, we derived the shape of the structure 
functions in the linear approximation. This shape can be compared with
the Batchelor's parametrization, which can be derived from the Kolmogorov
four-fifth law using matched asymptotics. This 
parametrization gives:
\EQ
S_2(\ell)=\frac{A (\ell/\eta)^2}
{\biggl[1+B(\ell/\eta)^2\biggr]^{(2-\zeta_2)/2}},
\label{batchelor}
\EN
where $A$ and $B$ are constants depending on the skewness of 
the velocity derivatives ($B\approx 7.2\times 10^{-3}$ \cite{stolo}). 
The best fit between Batchelor's parametrization and the generic 
solution is obtained for $\eta_2=6.35\eta$ and is shown in Fig. \ref{fig:three}. 

\subsubsection{General scaling}

\begin{figure}[!htb]
\includegraphics[width=0.99\columnwidth]{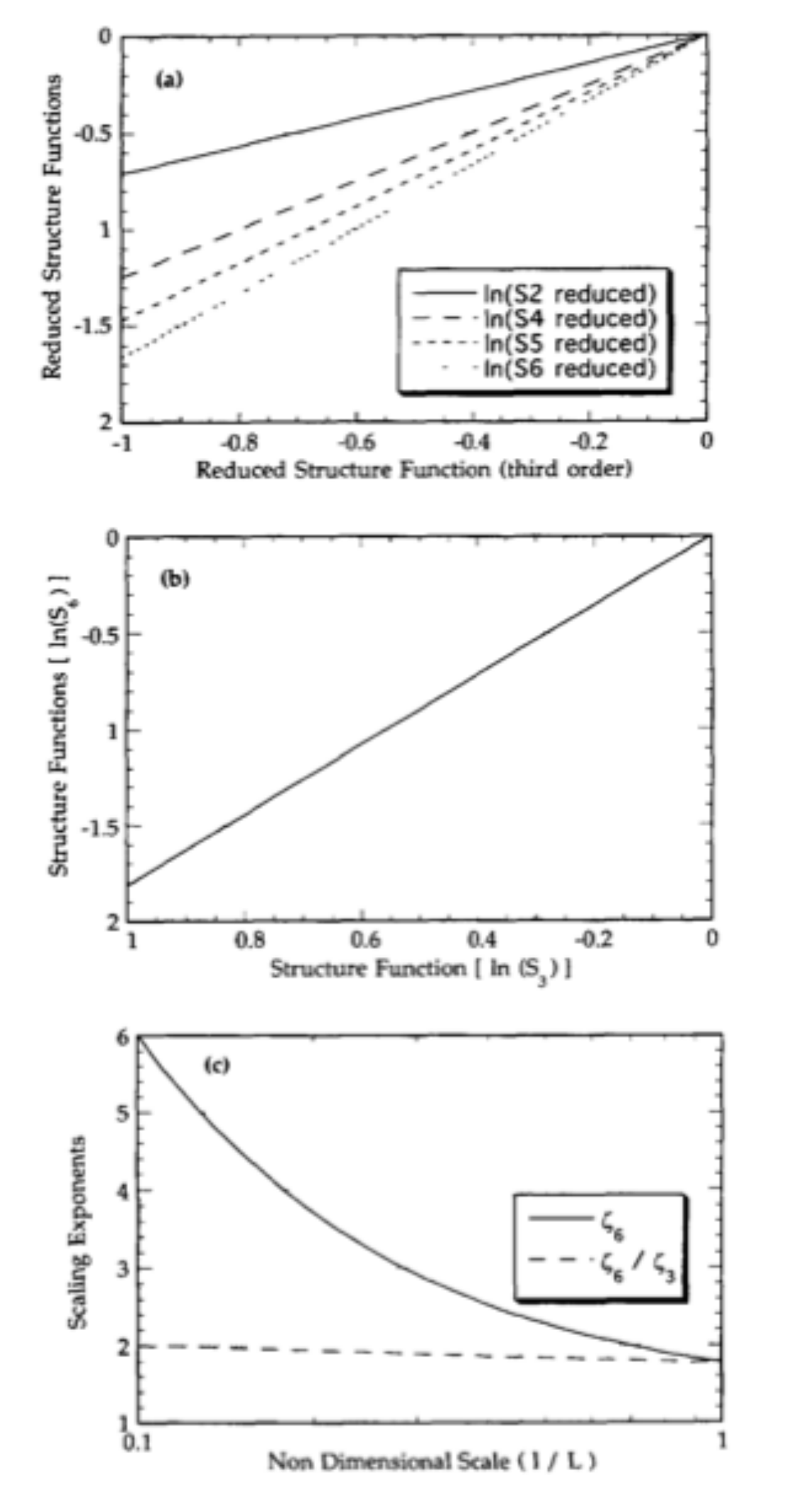}
\caption{Top: (a) General Scaling: all the reduced structure functions are proportional to
each other (here the third one).
Middle: (b) Extended Self Similarity: when the $6^{th}$ order structure
function is plotted against the $3^{rd}$ one, one observes a better
defined scaling regime (compare with Fig. 2). Despite the visual 
impression, the
Extended Self Similarity is only approximate (see Section 3).
Bottom: (c) Local scaling exponent $\zeta_6$ and relative scaling exponent
$\zeta_6/\zeta_3$ versus the scale. The local scaling exponent decreases
constantly with the scale, proving the absence of exact self-similarity.
By contrast, the relative scaling exponent varies much less, which 
enables a better determination of this exponent (ESS property).}
\label{fig:four}
\end{figure}

The structure functions computed in the previous Section display 
an interesting property, referred to in turbulence as 
General Scaling (GS) \cite{GS}. 
We introduce the maximal event function $S_\infty$ defined as:
\EQ
S_\infty=\lim_{n\to\infty}\frac{S_{n+1}}{S_n}.
\label{defsifty}
\EN
In a bounded system, this function is always defined and represents the 
event of maximal intensity. In turbulence, it represents the
largest velocity differences. 
The reduced structure functions $S_n/S_\infty^n$ then obey a remarkable
factorization property,
\EQ
\ln\biggl(\frac{S_n}{S_\infty^n}\biggr)\propto
\ln\biggl(\frac{S_3}{S_\infty^3}\biggr).
\label{facto}
\EN
This is shown in Fig. \ref{fig:four}a.
This factorization property, extending throughout the whole
scale interval, can be seen as a generalization 
of the factorization occurring in the inertial 
range, where all structure functions are proportional to $T$.\

This result can be explained by our choice of $n$ independent pseudo-Reynolds
numbers $R_n\equiv R$. 
In such case, from (\ref{solution}) and (\ref{interpretation}), we
obtain:
\EQ
\ln(S_\infty)= 
\Delta_\infty\frac{\ln(R)}{\ln(\Delta_\infty)}
\biggl(\Delta_\infty^{T/\ln(R) }-1\biggr).
\label{shapifty}
\EN
The reduced structure functions $S_n/S_\infty^n$ are then simply given by: 
\EQ
\ln\biggl(\frac{S_n}{S_\infty^n}\biggr)=
\frac{\delta\zeta_n}{1-\Delta_\infty}\biggl[T
-\Delta_{\infty}\frac{\ln(R)}{\ln(\Delta_\infty)}
\biggl(\Delta_\infty^{T/\ln(R) }-1)\biggr],
\label{factoexp}
\EN
and are proportional to each other within the whole scale interval.
 It is not clear whether the present analysis provides an explanation
to the GS observed in turbulence. 
It is not known in turbulence 
whether the pseudo-reynolds numbers $R_n$ are actually $n$ 
independent, i.e. if the transition from inertial range solution 
towards the regular solution occurs at the same scale $\eta_n=\eta$
(see \cite{GS} for a discussion about $n$ dependent cut-offs). 
If the dependence of $\ln(R_n)$ with $n$ is weak, as seems to be the 
case in turbulence, then the present analysis is still valid in turbulence
and explains the phenomenon of GS within the context of 
scale symmetry.

\subsubsection{Extended Self-Similarity}

The factorization property (\ref{facto}) can be strengthened if, in addition,
$\ln(S_\infty)$ is also proportional to any $\ln(S_n/S_\infty^n)$. In that
case \cite{GS}, the logarithms of any structure functions are proportional
to each other. In other words,
when one structure function
is plotted against another one (e.g. the third one), one observes a
well defined scaling regime, even in the range of scale 
where the function is not self-similar. This property was 
called Extended Self-Similarity \cite{ESS}. The
condition to observe this is $\Delta_\infty\to 0$.
In practise, if this condition is satisfied approximately, one
can expect to observe ESS in the system. In our case, $\Delta_\infty=1/9$
is small, so we should observe ESS.   
This is illustrated in 
Fig. \ref{fig:four}b. Note that ESS property means that 
the relative exponent 
$\zeta_n^\star$ defined as
\EQ
\zeta_n^\star=\frac{d\ln(S_n)}{d\ln(S_3)},
\label{defzetast}
\EN
is much better defined than the true exponent:
\EQ
\zeta_n=\frac{d\ln(S_n)}{d\ln(\ell)}.
\label{defzeta}
\EN
To illustrate this point, we have computed these scaling exponents for
$n=6$  (Fig. \ref{fig:four}c). It can 
be seen that $\zeta_6$ decreases steadily from $\zeta_6=6$ to
$\zeta_6=1.78$, while $\zeta_6^\ast$ displays much weaker variations 
over the whole interval, from $\zeta_6^\ast=2$ to
 $\zeta_6^\ast\approx 1.8$ (the ``intermittent value").\

\subsubsection{Finite size effects vs asymptotic K41 solution?}

In the previous Sections, we have used an interpretation of the constants 
to compute explicitly structure functions. This interpretation was dictated
by our choice to introduce explicitly the dissipative range in the boundary
condition (via the matching to the ``regular solution"). In such 
interpretation, asymptotic (high Reynolds number)
 scaling exponents take the shape (\ref{exponents}), with a linear part coming 
explicitly from finite size effects. In absence of finite size effects,
they take the simple shape $\zeta_n=C_{scale}(1-\beta^n)$ predicted by 
Dubrulle and Graner \cite{DG1}. The Kolmogorov solution 
$\zeta_n=n/3$ is never reached, unless the coefficient 
$\beta$ appearing in (\ref{solution}b) depends on other external parameters such
as the dimension. The Kolmogorov solution could then appear as the 
infinite dimension limit (even in presence of finite size effects), 
and $\beta$ would be a parameter characterizing
the dimension of the system.\

Other interpretations are however possible, within the same symmetry 
arguments. For exemple, one could restrict the solutions to an ``inertial
range" of scale defined by imposing $\zeta_3=1$, i.e. $\alpha_3=0$ in
(\ref{poisson}). Asymptotic Kolmogorov solution could then be obtained 
with the following choice of constants (still compatible with 
the theory of Dubrulle-Graner):
\EQA 
\frac{\chi_n}{a_n}&\equiv&\frac{\delta\zeta_n}{\delta\zeta_3},\nonumber\\
-\alpha_n\frac{a_n}{b_n\ln(R_n)}&
\equiv&\frac{n}{3}-\frac{\delta\zeta_n}{\delta\zeta_3},\nonumber\\
e^{-a_n/b_n}&=&3\Delta_\infty.
\label{newinter}
\ENA
In such case, the derivative $d\ln(S_n)/dT$ reaches the value (\ref{dgpred}) 
at $T=\ln(R_n)$, which appears as another boundary condition. In absence 
of finite size effects, the scaling exponents are 
$\zeta_n=(1-\beta^n)/(1-\beta^3)$. The Kolmogorov solution $\zeta_n/3$ 
is obtained
as the asymptotic solution (Reynolds tends to infinity)
 in presence of finite size effects.\  

We were not able to discriminate between these two types of interpretations.
Obviously, one is rather valid in the vicinity of the dissipative range.
It could then be seen as a refinement of the arguments by Frisch and Vergassola
\cite{FV} obtained within the multifractal model (see \cite{castaingetal}
for a discussion within the log-similarity hypothesis). The second
is valid in the ``inertial range", defined using $\zeta_3=1$ (inertial
range log-similarity hypothesis, see \cite{castaingetal}). They 
however lead to
distinct prediction about the possibility to approach Kolmogorov solution. 
The most recent experimental results seem to indicate that the observed
scaling exponents are almost independent of the Reynolds number \cite{arneodo}.
This would favor the first interpretation, and explain our choice in 
the present paper. Obviously, it would be interesting to consider 
further the variation of scaling exponents with the Reynolds 
number, and, possibly, to reconsider the second interpretation.\

It should be stressed however that in any case, the ``exponential 
power-law" appears as a prediction of symmetry arguments, independent 
of any boundary consideration, i.e. of these kind of interpretations.

\section{Discussion}

Using only symmetry considerations, we were able to build generic
structure functions reproducing many features observed in 
actual structure functions in turbulence: transition from exponential
power shape to a power shape with increasing Reynolds number, 
extended-self-similarity, regular matching with the regular solution
at small scale. We do not claim that the generic structure functions
considered in the present paper actually fit exactly the structure
functions determined experimentally, because we intentionally considered
only the simplest case relevant to turbulence, where the equation 
is linear. It is clear for example that the observed
 large scale saturation 
of the structure functions, which is absent in the linear model, 
could be obtained by taking into account non-linear terms in the 
differential equation, i.e. by a slightly more complicated model.\

Our results illustrate the essential influence of scale symmetry 
on structure functions in turbulence, and provides further 
support to the scenario of "scale invariant" anomalous scaling
discussed in Section 2. We note that finite size effects 
(ultraviolet cut-off) generically lead to non-power law behavior
of the structure function, but rather to exponential power-law behavior, 
$\exp(\ell^\alpha)$, where $\alpha$ is real, proportional to the 
inverse Reynolds number. Such dependence, connected with the 
requirement of covariance by resolution, was also
inferred by Barenblatt and Goldenfeld
using the principle of "Reynolds number covariance" \cite{barenblatt2}.
We see here that such principle (left unjustified by 
Barenblatt and Goldenfeld) is directly connected with scale symmetry.
Finally, we note that there is a possibility to get complex 
exponential power-law behaviors ($\alpha$ complex) if we allow 
a differential equation of higher order, or if we allow the presence 
of terms directly proportional to $A$ in (\ref{ampli}). The first 
possibility could be justified if more than two boundary conditions
are necessary to specify the solution. The second possibility
still requires only two boundary conditions, but implies a breaking 
of global scale invariance. This requires the existence of
a privileged scale into the system, and occurs for example 
in a system subject only to discrete scale invariance \cite{sornette}.
Complex exponents gives rise to log-periodic oscillations at 
large Reynolds number (when the solution goes from exponential
power law to power law), which may have been detected in a variety 
of physical systems \cite{sornette,TB}. It would be interesting
to see whether they can also arise in certain turbulent flows.

\par {\bf Acknowledgements}\
 
We thank F.M  Br\'eon for assistance in the preparation of the manuscript
and M. Vergassola and G. He for useful comments.
This work was supported by a grant from
the french Caisse d'Allocations Familiales. \


\begin{thebibliography}{99}
\bibitem{remarke}
Note that until now, the explicit shape of the operator $I_n$
has never been computed, even perturbatively.
\bibitem{remark0}
Because if the global scale invariance, it does not matter 
whether we work with $\ln\ell$ or with $\ln(\ell/\ell_0)$, where
$\ell_0$ is any characteristic scale.
\bibitem{remark2}
 Note that a linear
equation in the variable $A$ is not necessarily linear in the 
variable $S_n$.
\end{thebibliography}
\end{document}